\documentclass[12pt]{article}
\usepackage{authblk}
\usepackage[margin=1in]{geometry}
\usepackage{amsmath, amssymb}
\usepackage{graphicx}
\usepackage[hyperfootnotes=false]{hyperref}
\usepackage{natbib}
\usepackage[version=4]{mhchem}

\begin{document}

\title{Phase behavior of thermoresponsive colloids drives re-entrant plasmon coupling}
\date{}

\author[a,b]{Angela Capocefalo}
\author[a,b,$\ast$]{Francesco Brasili}
\author[c]{Javier P\'erez}
\author[d]{Edouard Chauveau}
\author[e]{Stefano Casciardi}
\author[e]{Andrea Militello}
\author[b]{Francesco Sciortino}
\author[a,b]{Emanuela Zaccarelli}
\author[b]{Federico Bordi}
\author[d,$\dagger$,$\ast\ast$]{Domenico Truzzolillo}
\author[a,b,$\dagger$]{Simona Sennato}

\affil[a]{\textit{Institute for Complex Systems, National Research Council,\newline Piazzale Aldo Moro 5, 00185, Roma, Italy}}
\affil[b]{\textit{Physics Department, Sapienza University of Rome,\newline Piazzale Aldo Moro 5, 00185, Roma, Italy}}
\affil[c]{\textit{Synchrotron SOLEIL, L'Orme des Merisiers,\newline D\'epartementale 128, 91190, Saint-Aubin, France}}
\affil[d]{\textit{Laboratoire Charles Coulomb, CNRS--Universit\'e de Montpellier,\newline UMR 5221, 34095, Montpellier, France}}
\affil[e]{\textit{Dipartimento di medicina, epidemiologia, igiene del lavoro e ambientale, INAIL,\newline Via Fontana Candida 1, 00078, Monteporzio Catone, Roma, Italy}}

\maketitle
\begingroup
\renewcommand{\thefootnote}{$\ast$}
\footnotetext{\texttt{francesco.brasili@cnr.it}}
\renewcommand{\thefootnote}{$\ast\ast$}
\footnotetext{\texttt{domenico.truzzolillo@umontpellier.fr}}
\renewcommand{\thefootnote}{$\dagger$}
\footnotetext{D.T.\ and S.S.\ contributed equally to this work.}
\endgroup

\vspace{-8mm}
\begin{abstract}
Plasmonic nanoparticles (NPs) integrated within thermoresponsive polymeric microgels provide a versatile platform for the realization of stimuli-responsive optical materials, where the microgel volume phase transition enables dynamic control of plasmon coupling. This study uncovers a counter-intuitive re-entrant behavior with increasing NP loading in which plasmon coupling initially strengthens and subsequently weakens beyond a critical NP-to-microgel number ratio. By combining light and X-ray scattering techniques with optical spectroscopy and electrophoretic mobility measurements, it is demonstrated that plasmon coupling is governed not only by the interparticle distance between NPs confined within individual microgels, but also by the colloidal stability of the hybrid complexes. At intermediate NP loadings, surface charge inhomogeneities induced by NP adsorption promote aggregation of microgel-NPs complexes, resulting in enhanced plasmon coupling. In contrast, when the complexes remain colloidally stable, coupling is dictated solely by NP organization within the corona of individual microgels. A quantitative relationship between plasmon coupling and interparticle distance reveals two distinct coupling regimes. This behavior is rationalized through a phase diagram linking colloidal stability to optical response. These findings identify colloidal stability as a key parameter for designing soft plasmonic systems with programmable optical properties.
\end{abstract}

\section{Introduction}
The integration of plasmonic nanoparticles (NPs) with soft polymeric colloids, such as microgels, has emerged as a convenient strategy for designing smart photonic materials with tunable optical properties ~\cite{karg2009,arif2021,diehl2022}. The unique and programmable features of these systems arise from the interplay between the dual colloidal-polymeric nature of  microgels \cite{fernandez2011,suzuki2023}, which enables precise control over softness and responsiveness to external stimuli, and the distinctive optical properties of the NPs~\cite{halas2011,amendola2017,wang2020}. In particular, noble metal NPs exhibit resonant interaction with visible light, at the so-called localized surface plasmon resonance (LSPR), whose spectral features are highly sensitive to interparticle spacing and organization through plasmon coupling effects, which occur when interparticle gaps fall within a few nanometers~\cite{biswasray2025}.

Thermoresponsive microgels offer a versatile platform to reversibly modulate NP organization via their volume phase transition (VPT), i.e. the collapse of the polymer network upon increasing temperature. The VPT has been widely exploited as a mechanism to bring NPs in close proximity and thereby control plasmon coupling~\cite{gorelikov2004,lange2012,gawlitza2013,honold2015,ye2024,brasili2025}, establishing microgel–NPs complexes as attractive platforms for sensing, photothermal therapy or nanocatalysis~\cite{choe2018,sershen2000,das2007,chang2023,sajid2025}. These features place microgel–NPs complexes within the broader context of soft matter photonics, which exploits the intrinsic flexibility of soft materials to control light–matter interactions~\cite{chen2020,ma2025}. Such hybrid systems provide novel routes for manipulating light at sub-diffraction length scales, highlighting the importance of understanding how their properties can be optimized to achieve the desired optical response.

The simplest and one of the most commonly used strategies to incorporate NPs within microgels relies on electrostatic attraction between the NPs and oppositely charged groups within the polymer network~\cite{das2007,regmi2010,davies2010,bradley2011}. Once adsorbed, stronger short-range interaction between the metal surface and the polymer could help anchor and stabilize the NPs~\cite{gawlitza2013}. We recently investigated in detail these electrostatic interactions between thermoresponsive poly(N-isopropylacrylamide) (pNIPAM) microgels and gold NPs, addressing both the impact of NP adsorption on the structure of the polymer network~\cite{brasili2023}, as well as the spatial organization of NPs within the spherical corona of the microgel and their rearrangement driven by the microgel VPT~\cite{brasili2025}. We also established a quantitative relationship between plasmon coupling and thermally-tunable NP spacing. However, the non-uniform surface charge distribution, resulting from NP adsorption, can lead to unexpected aggregation phenomena, including finite-size clustering and overcharging, \textit{i.e.}, the reversal of the sign of the net charge of the complexes compared to bare microgels, which results in a re-entrant condensation~\cite{bordi2009}. This non-trivial phenomenology has been previously observed separately in both microgel-based systems, where thermoresponsive microgels were decorated by smaller species~\cite{truzzolillo2018,sennato2021}, and plasmonic NPs decorated by globular proteins~\cite{brasili2020,capocefalo2022}. Instead, the role of colloidal stability in determining the optical properties of microgel–NPs complexes has remained largely unexplored. Previous studies have primarily focused on tuning plasmon coupling by varying the microgel crosslinker density or the NP surface coverage~\cite{lange2012,gawlitza2013}, while largely overlooking the colloidal behavior of the resulting hybrid complexes. 

In this work, we tackle this problem by systematically analyzing the optical properties of microgel–NPs complexes across the VPT, while varying the NP-to-microgel number ratio $n$. We reveal a counter-intuitive modulation of plasmon coupling, which initially strengthens with increasing $n$ and then weakens again beyond a critical threshold. To rationalize these findings, we report a small angle X-ray scattering (SAXS) analysis of the NP--NP correlation to establish a direct, quantitative connection between NP gathering across VPT and the resulting optical response of the system in terms of plasmon coupling. We find a dual behavior depending on $n$, that we finally explain within the framework of finite-size clustering and re-entrant condensation of charge-patched colloids, through hydrodynamic radius, electrophoretic mobility and electron microscopy measurements.

We unveil a re-entrant plasmon coupling in the microgel–NPs system that is governed primarily by complexes aggregation, in addition to the number and relative positioning of NPs adsorbed within the microgel corona. By identifying the key parameters controlling plasmon coupling in dependence of the colloidal properties of the system, this work establishes a rational framework for tuning the optical properties of these hybrid plasmonic systems, thus paving the way toward the design of responsive plasmonic materials with programmable optical properties for sensing, photonic, and nanomedicine-related applications.

\section{Results and discussion}
\subsection{Thermal modulation of the optical response}
\begin{figure}[b]
\centering
\includegraphics[width=\linewidth]{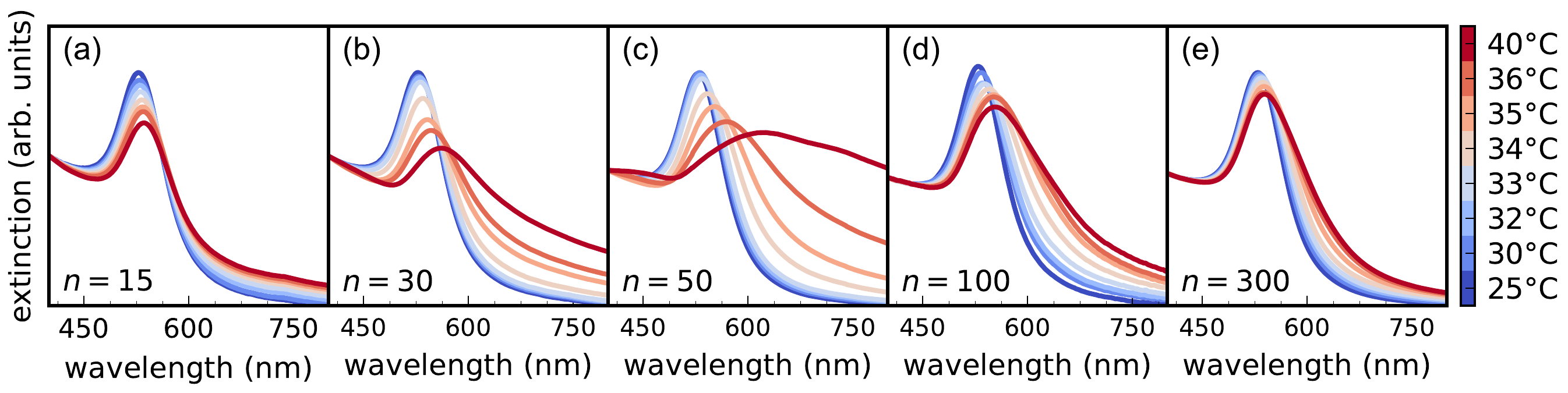}
\caption{Extinction spectra of microgel-NPs as a function of temperature for selected values of the NP-to-microgel number ratio $n$: 15 (a), 30 (b), 50 (c), 100 (d), and 300 (e).}\label{fig:fig1_spectra}
\end{figure}

\begin{figure}[t!]
\centering
\includegraphics[width=0.95\linewidth]{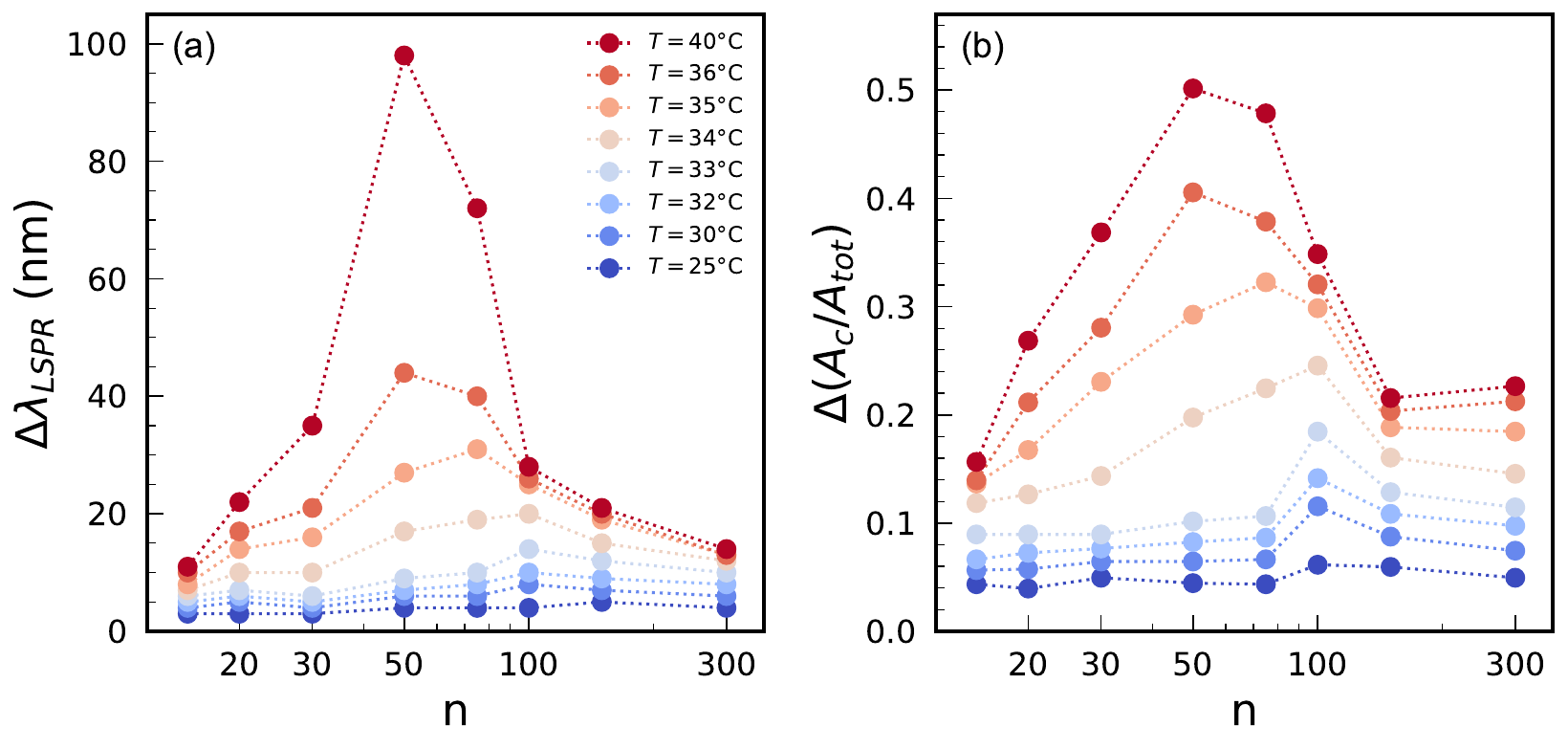}
\caption{(a) Shift of the LSPR peak $\Delta\lambda_\text{LSPR}$ and (b) degree of plasmon coupling $\Delta (A_C/A_{tot})$, defined as the spectral weight of the region of coupled plasmon modes ($\lambda>570$ nm, Equation~\ref{eq:coupling}), as a function of $n$ for different $T$. }\label{fig:fig2_coupling}
\end{figure}

We first investigate the thermal modulation of the optical properties of microgel-NPs complexes with different NP-to-microgel number ratios $n$ by extinction spectroscopy. The complexes are obtained by electrostatic adsorption of anionic gold NPs, with radius $a=9.3$ nm and LSPR peak at $\lambda_\text{NP}=525$ nm, to cationic microgels with VPT at $T_\text{VPT}=32$°C. Control spectra of stock NPs and pristine microgels dispersions, reported in Figure~\ref{fig:figS1_refs}, demonstrate that the contribution to the extinction due to scattering from microgels \cite{otten2024} is negligible compared to the LSPR of NPs, even in the collapsed state. The spectra for selected values of $n$ are shown in in Figure~\ref{fig:fig1_spectra}, additional datasets are reported in Figure~\ref{fig:figS2_spectra}. At 25°C, below $T_\text{VPT}$, no signs of plasmon coupling can be detected, as all LSPR peaks exhibit the typical spectral shape of isolated, non-interacting NPs. As temperature increases, the spectra evolve, exhibiting both a redshift of the LSPR peak, and an increase in optical extinction at longer wavelengths ($\lambda \ge 570$ nm).
These spectral changes are characteristic of plasmon coupling~\cite{halas2011}, therefore one would be tempted to attribute them to the progressive gathering of adsorbed NPs, driven by the shrinking of microgels across the VPT. Under this interpretation, the coupling would be expected to become progressively stronger with increasing $n$, as the reduced surface-to-surface distance $d$ for higher NP loading (growing with $n$~\cite{brasili2025}) would result in stronger plasmon coupling. Instead, the spectral features of plasmon coupling appear to be most evident at intermediate coverage ($20\lesssim n\lesssim 100$), and become less pronounced for the highest $n$ values. We take a deeper look at this unexpected behavior by reporting the LSPR shift $\Delta\lambda_\text{LSPR}$ with respect to $\lambda_\text{NP}$, and the degree of plasmon coupling $\Delta (A_C/A_{tot})$, defined in Equation~\ref{eq:coupling}, as a function of $n$ for all the analyzed temperatures in Figure~\ref{fig:fig2_coupling}a and b, respectively. Both quantities increase with $T$ at fixed $n$, while, at fixed $T$, they display a non-monotonic dependence on $n$. In particular, for temperatures above the VPT, they increase up to $n \simeq 50$ and then decrease back for higher values of $n$. Below the VPT, this effect is shifted to larger $n$ and barely noticeable.

These results demonstrate a re-entrant behavior of the optical properties, which has not been previously observed and thus deserves a dedicated investigation to provide a proper interpretation and identify the underlying mechanisms of plasmon coupling. In this respect, while $\Delta\lambda_\text{LSPR}$ is useful for applications such as colorimetric sensing, it has two key limitations, being sensitive to changes in the dielectric environment caused by water expulsion across the VPT, and failing to fully capture the emergence of coupled plasmon modes at longer wavelengths. In contrast, the degree of coupling accounts for the entire spectral modifications and thus better represents the plasmon hybridization that occurs upon NP approach. In the following section, we therefore combine SAXS measurements with optical analysis to establish a quantitative relationship between plasmon coupling and NP structural arrangement within the microgels, using $\Delta (A_C/A_{tot})$ as the key optical parameter.

\subsection{Linking optical properties to NP structure}
\begin{figure}[b!]
\centering
\includegraphics[width=\linewidth]{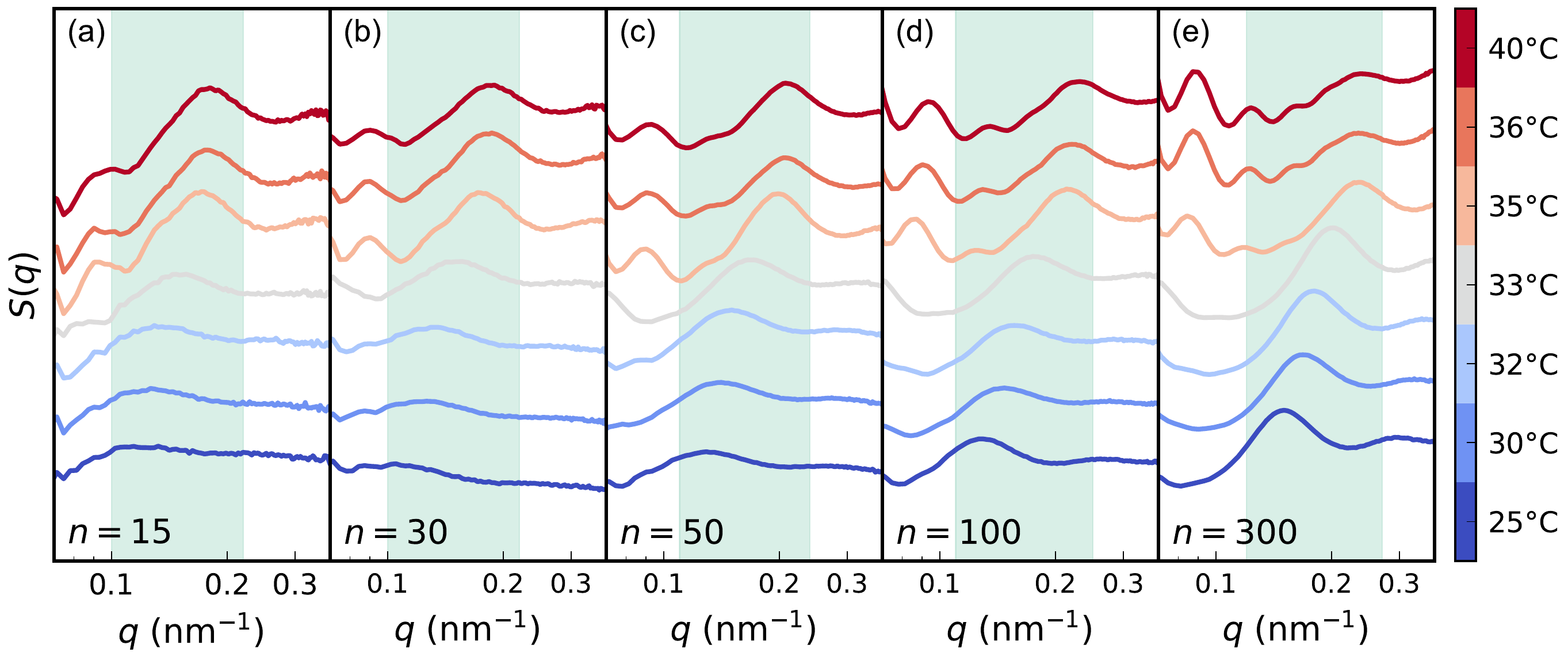}
\caption{Structure factors of the microgel-NPs samples as a function of temperature for the same values of $n$ of Figure~\ref{fig:fig1_spectra}. The light green color highlights the peaks corresponding to the distance between NPs.}\label{fig:fig3_saxs}
\end{figure}

\begin{figure}[t]
\centering
\includegraphics[width=\linewidth]{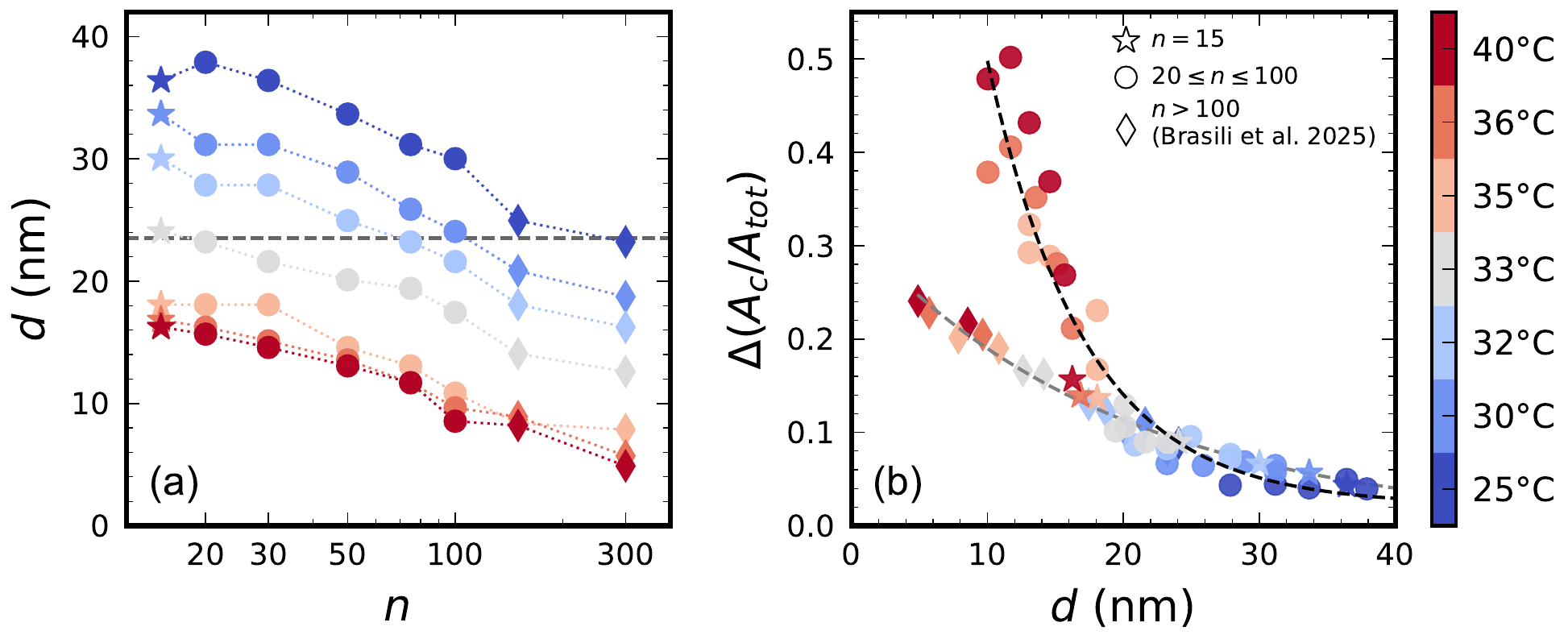}
\caption{a) Surface-to-surface distance $d$ between NPs as a function of $n$ for different temperatures, derived from structure factors of NPs incorporated in the microgel corona; the horizontal dashed line ($d=23.7$ nm) marks the intersection between the exponential fits in panel b. b) Degree of plasmon coupling $\Delta (A_C/A_{tot})$ as a function of $d$; dashed lines are the fits to exponential decay.}  \label{fig:fig4_linking}
\end{figure}

To gain knowledge on the microscopic organization of NPs adsorbed to microgels, we study the thermal evolution of their structure factors $S(q)$ for different $n$. Measurements are performed by SAXS, thus ensuring that the measured intensity originates exclusively from the NPs, thanks to their much higher contrast compared to the polymer~\cite{suzuki2014}. The corresponding $S(q)$ are shown in Figure~\ref{fig:fig3_saxs} for the same values of $n$ selected for Figure~\ref{fig:fig1_spectra}, additional datasets are reported in Figure~\ref{fig:figS3_saxs}. The structure factors exhibit several features resulting from the complex structure of the NPs embedded within microgels, whose thermal evolution has a non-trivial interpretation that we provided in a previous work with the aid of numerical simulations~\cite{brasili2025}. This interpretation involves the formation of a shell of NPs adsorbed in the outer corona of the microgel, which becomes more compact and narrow, driven by the VPT of the microgel. In this context, it is possible to identify two distinct $q$-regions, approximately below and above 0.1 nm$^{-1}$, associated with structural correlations at different length scales. Here, we are interested in the surface-to-surface distance $d$ between nearest neighbor NPs as a function of temperature, therefore we focus on the position $q_p$ of the first peak in the high-$q$ range, following the methodology of our previous study. The evolution of $q_p$ - approximately between 0.1 and 0.3 nm$^{-1}$, depending on $n$ - is highlighted in Figure~\ref{fig:fig3_saxs} by light green regions. The average surface-to-surface distance between NPs $d(n,T)$, given by Equation~\ref{eq:interparticle_distance} for each experimental condition ($n,T$), is reported in Figure~\ref{fig:fig4_linking}a. As expected, the data show that $d$ decreases as temperature increases for all the investigated values of $n$, reflecting NP gathering driven by the VPT. At fixed $T$, $d$ also decreases with increasing $n$, consistent with the larger number of NPs adsorbed to the microgels. However, the data do not appear to correlate with the re-entrant behavior observed in the optical properties. To explore this aspect, we plot $\Delta (A_C/A_{tot})$ as a function of $d$ in Figure~\ref{fig:fig4_linking}b. At large $d$ values, the data collapse onto a single master curve. As $d$ decreases, they split into two distinct regimes. Surprisingly, the same interparticle separations correspond to markedly different coupling strengths. In the weaker-coupling branch, the data for $n=15$ are in excellent agreement with those for $n>100$ reported in ref.~\citenum{brasili2025}, whereas all intermediate $n$ data fall on a second branch exhibiting significantly stronger coupling.

Following the model in ref.~\citenum{jain2007}, we fit separately each branch with an exponential decay $\Delta (A_C/A_{tot}) = C\exp(-d/l_d)$, where $l_d$ is the characteristic decay length, quantifying the attenuation of the near field coupling, and $C$ represents an effective coupling strength. We find $C=0.3$ and $l_d=19.1$ nm for the lower branch, while $C=1.9$ and $l_d=7.2$ nm for the higher one, revealing the onset of two distinct mechanisms governing plasmon coupling. The second one, that is prevalent at intermediate $n$, leads to stronger coupling but at the same time has a more rapid decay as a function of $d$. In this respect, we must consider that plasmon coupling in NP assemblies also depends on other parameters than the interparticle distance $d$, including the overall size of the assemblies and their spatial arrangement, which can vary from compact to highly ramified structures and is often characterized in terms of branching and fractal dimension~\cite{capocefalo2022,esteban2012,taylor2011}. These features can be affected by the aggregation of microgel–NP complexes, that would bring NPs confined to different microgels in close proximity, resulting in larger NP clusters. However, $d$ represents the nanoscale distance between nearest-neighbor NPs, without distinguishing whether they belong to the same microgel or two approaching microgels. Assessing the colloidal stability of microgel-NP complexes requires probing larger length scales (microscopic range), which are not fully accessible in our SAXS measurements.
Therefore, in the following paragraph, we report a DLS investigation of the microgel-NPs complexes aimed at constructing their phase diagram to gain further insight on the two coupling mechanisms and to link them to the re-entrant behavior of plasmon coupling.

\subsection{Re-entrant condensation of microgel-NPs complexes drives plasmon coupling}
We interpret the re-entrant optical behavior of the microgel-NPs complexes within the context of the colloidal stability of charged particles co-dispersed with oppositely charged polyelectrolytes. The electrostatic adsorption of polyelectrolytes leads to heterogeneous charge distribution on the primary particle's surface, resulting in charge-patched colloids, and introducing an attractive contribution to the effective interparticle interactions. This affects the colloidal stability and promotes the formation of stable aggregates, whose finite size increases with the concentration of the adsorbing species~\cite{bordi2009}. Moreover, due to the strong electrostatic correlations between the adsorbed polyelectrolytes and the charge fractionalization mechanism \cite{grosberg2002}, it is possible that polyelectrolytes adsorb in excess with respect to the amount required for complete neutralization of the surface charge, resulting in overcharging, i.e., the inversion of the net charge sign of the complex relative to the primary particles. As a consequence, re-entrant condensation occurs: beyond a certain polyelectrolyte loading the aggregate size reaches a maximum and then decreases upon further adsorption.
This phenomenology has been extensively studied in systems where liposomes~\cite{bordi2009,truzzolillo2015} or plasmonic NPs~\cite{brasili2020,capocefalo2022} act as primary colloids, and has more recently been reported also for microgels~\cite{truzzolillo2018,sennato2021}. In particular, in the case of thermoresponsive microgels, aggregation phenomena can be further modulated by temperature, as the collapse of the polymer network across the VPT leads to a substantial increase in their surface charge density. Consequently, surface charge heterogeneities, responsible for the attractive contribution in the pair interaction potential, can increase significantly. 
As a result, the effective interparticle interactions become strongly temperature-dependent, adding an additional level of complexity to the aggregation behavior. This effect has been reported in ref.~\citenum{sennato2021}, where a similar phenomenology was observed when the adsorbing species were silica NPs rather than charged macromolecules. In the case of plasmonic NPs, such temperature-driven modulation of patchy interactions is expected to introduce new interaction regimes, with direct consequences on NP organization and on the resulting plasmon coupling.
\begin{figure}[t]
\centering
\includegraphics[width=\linewidth]{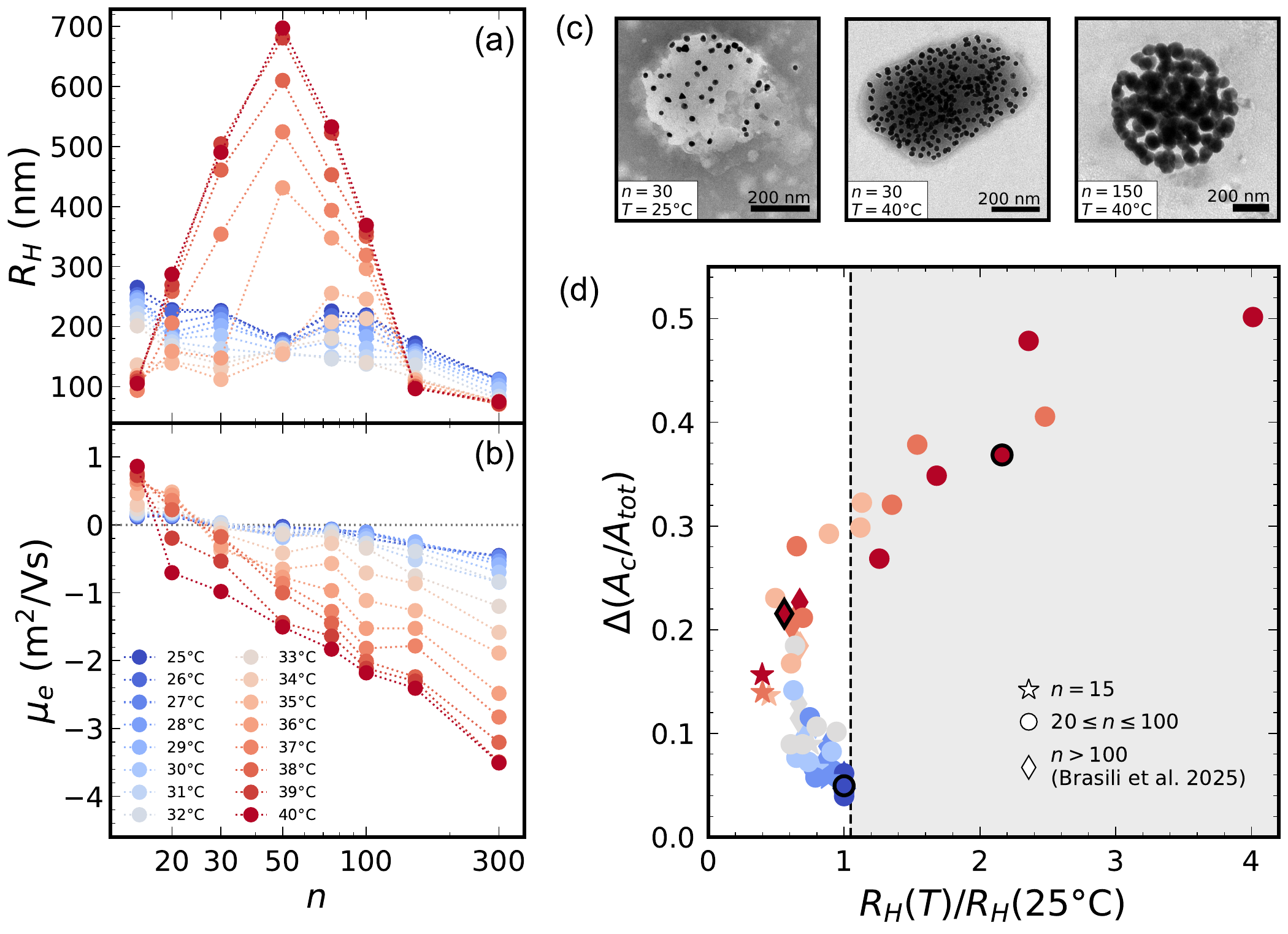}\caption{Hydrodynamic radius $R_H$ (a) and electrophoretic mobility $\mu_e$ (b) of the microgel-NPs complexes as a function of $n$, for different temperatures across the VPT. (c) Representative electron microscopy images of the samples in selected ($n$,$T$) conditions: individual complexes for $(30,25$°C$)$ and $(150,40$°C$)$, and aggregated structures, for $(30,40$°C$)$. (d) Degree of plasmon coupling $\Delta (A_C/A_{tot})$ as a function of $R_H(n,T)/R_H(n,25$°C$)$; the vertical dashed line divides the region of colloidal stability from that of complex aggregation (shaded in gray); the data points highlighted by black edges correspond to the images in panel c.} \label{fig:fig5_dls}
\end{figure}

Here we analyze the thermally-induced aggregation of microgel-NPs systems for different values of $n$, and relate it to their optical properties. In Figure \ref{fig:fig5_dls}a and b, we report the trends of hydrodynamic radius $R_H$ and electrophoretic mobility $\mu_e$ as a function of $n$ for a range of temperatures across $T_\text{VPT}$. Below $T_\text{VPT}$, the values of $R_H$ are always smaller than those of bare microgels, demonstrating the colloidal stability of the complexes over the entire range of $n$ investigated. Moreover, the values of $R_H$ decrease with $n$, particularly for $n=15$, consistent with the NP-induced shrinking of the microgels~\cite{brasili2023}. As temperature increases, the values of $R_H$ decrease, following the VPT, until, for some intermediate values of $n$ ($20\leq n\leq 100$), they abruptly increase, becoming larger than those of individual complexes at 25°C. This indicates the formation of aggregates, consistent with the charge-patch aggregation phenomenon. The size of the aggregates increases with $n$ up to a maximum around $n=50$, and then decreases until the colloidal stability of the primary colloids (microgel-NPs complexes) is restored for $n>100$. Examining the behavior of $\mu_e$, we observe that the microgels are initially positively charged due to the amine groups from the initiator of the chemical synthesis. The charge variations become more evident at higher temperatures because the surface charge density grows with the compaction of the surface area and because mobility is enhanced by the reduced particle size. 
As $n$ increases, the adsorption of anionic NPs lowers the net charge of the complexes until charge inversion occurs, between $n=20$ and $n=30$, depending on temperature. Beyond the isoelectric point the surface charge density becomes progressively more negative and inhomogeneous, and this is what drives the re-entrant condensation. Importantly, throughout this process the surface coverage by adsorbed NPs remains low, even in the collapsed state ($\theta<25\%$ for $n\leq100$, see Section \ref{sec:S2_overcharging} and Figure \ref{fig:fig5_dls}c), so that the microgel surface never becomes uniformly charged and the charge-patch description holds over the entire range explored. Considering the surface coverage at charge inversion, a simple geometrical argument shows that the NP-covered regions are far more densely charged than the bare microgel surface. This strong charge asymmetry explains why the maximum aggregate size occurs at a loading well beyond that of net-charge neutrality (see Section \ref{sec:S2_overcharging}).

It is worth noting that thermally-induced aggregation occurs over the same range of $n$ corresponding to the stronger-coupling branch in Figure \ref{fig:fig4_linking}b. To quantitatively investigate this aspect, we plot the plasmon coupling obtained from extinction spectra in different experimental conditions $(n,T)$ as a function of the ratio $R_H(n,T)/R_H (n,25$°C$)$ measured under the same conditions (Figure \ref{fig:fig5_dls}d). We find that all data collapse onto a common trend, with $\Delta(A_C/A_{tot})$ increasing both when the microgel VPT induces particle shrinking without compromising colloidal stability ($R_H(n,T)/R_H (n,25$°C$)\leq1$), and, even more markedly, when the complexes aggregate (gray-shaded region). 
One might wonder whether the growth of extinction at $\lambda>570$ nm, and hence the re-entrant behaviour of $\Delta(A_C/A_{tot})$, could be attributed to scattering from the large aggregates that form under the same experimental conditions. This hypothesis is ruled out qualitatively by the concurrent redshift of $\lambda_\text{LSPR}$, which mirrors the behaviour of $\Delta(A_C/A_{tot})$ (Fig.~\ref{fig:fig2_coupling}): a featureless scattering background cannot produce such redshift, which is instead a specific signature of interparticle plasmon coupling. We further substantiate this conclusion quantitatively in Section~\ref{sec:S3_corrections}, where we correct the extinction spectra by subtracting the contribution of size-matched NP-free microgel aggregates. This analysis shows that aggregate scattering leads at most to a slight underestimate of the coupling degree, leaving the re-entrant dependence on $n$ and the overall picture presented here unaffected.

These results provide the first direct evidence of a relationship between the plasmon coupling of NPs adsorbed to thermoresponsive microgels and the colloidal stability of the microgel-NPs complexes. Moreover, by using the same markers of Figure \ref{fig:fig4_linking}b, we set a correspondence with the two branches of that plot. It then becomes evident that the data points belonging to the strong-coupling branch coincide with the onset of aggregation, demonstrating that the formation of NP assemblies extended across the corona of approaching microgels is responsible for the enhanced plasmon coupling observed, beyond that induced by temperature increasing in isolated complexes. Indeed, in the case of stable suspensions, plasmon coupling originates from interactions among neighboring NPs confined within the spherical corona of individual microgels~\cite{brasili2025}. In contrast, upon aggregation of the microgel-NPs complexes, the NP assemblies responsible for coupling extend across the coronas of two or more adjacent microgels, thereby involving a larger number of NPs and effectively generating two interacting NP layers at each contact surface. This configuration, reminiscent of that reported for superimposed planar NP layers~\cite{fasolato2014}, results in a substantially amplified coupling strength. The electron microscopy images of Figure \ref{fig:fig5_dls}c further support this interpretation, providing direct visualization of isolated complexes, for the samples $(30,25$°C$)$ and $(150,40$°C$)$, and aggregated structures, for the sample $(30,40$°C$)$, where NPs can be clearly observed in contact across the coronas of neighboring microgels.
\begin{figure}[t]
\centering
\includegraphics[width=0.6\linewidth]{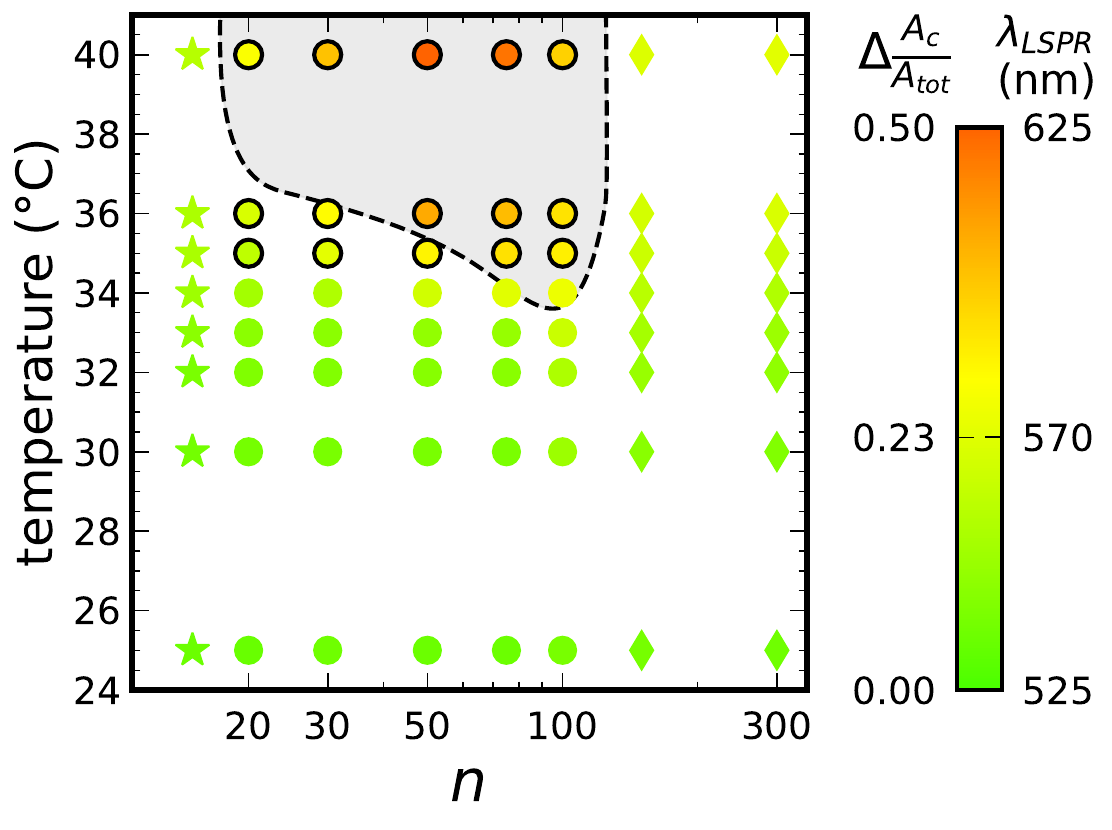}\caption{Phase diagram of microgel-NPs in the $n-T$ plane: the dashed black line delineates the boundary between colloidal stability of the complexes and the aggregation region (shaded in gray), as identified from DLS measurements by applying the threshold $R_H(n,T)>R_H(n,25\text{°C})$, to define the aggregation onset; the colored points denote the samples for which extinction spectra were collected: we used the same markers of Figure \ref{fig:fig4_linking}b to set a correspondence with the two branches of that plot (circles with black edges identify the high-coupling branch), while the color is assigned based on values of $\Delta(A_C/A_{tot})$, according to the colormap on the right; the colormap encodes the LSPR wavelength extracted from the extinction spectra, providing a direct visualization of the optical properties of the system.} \label{fig:fig6_phasediagram}
\end{figure}

Building on this evidence, we establish a direct correspondence between the phase diagram of the colloidal dispersion and the optical response of the system. In Figure \ref{fig:fig6_phasediagram}, we delineate the regions of the $n-T$ plane where the complexes are colloidally stable from those where they aggregate (grey area). The aggregation region coincides with the regime of strongest coupling identified in Figure \ref{fig:fig4_linking}b for intermediate $n$. Moreover, the shorter decay length found for aggregating complexes is consistent with previous results obtained for interacting planar layers of NPs \cite{ung2001}. Each experimental condition $(n, T)$ investigated by extinction spectroscopy is then mapped onto the same phase diagram. We assigned colors based on the magnitude of $\Delta(A_C/A_{tot})$ on the right scale, that is mapped on the corresponding LSPR peak wavelength $\lambda_\text{LSPR}$, assigning to each wavelength its corresponding spectral color. We further highlight the points corresponding to the stronger-coupling branch using circles with a black edge. This diagram clearly shows that the aggregation of the complexes is precisely what determines a more intense plasmon coupling compared to that observed as a function of temperature for individual complexes. The onset of aggregation thus marks a distinct optical regime characterized by a pronounced enhancement of plasmon coupling besides that observed for isolated complexes upon temperature increase.

We finally note that the re-entrant plasmonic response reported here is not fully reversible under heating–cooling cycle: for samples that aggregate, both colloidal structure and optical properties persist in the high-coupling state upon cooling; for non-aggregating samples, the VPT is fully reversible and the optical response is largely recovered, with small residual offset at 25°C. A detailed analysis of the thermal reversibility is reported in Section~\ref{sec:S4_reversibility}.

Overall, these findings establish a unified framework linking thermally driven phase behavior of microgel-NPs complexes to the collective plasmonic response, providing a powerful approach to rationally design responsive plasmonic materials through controlled modulation of colloidal interactions.

\section{Conclusions}
Microgel–NPs complexes represent a versatile class of plasmonic colloidal systems in which plasmon coupling can be dynamically modulated by exploiting the intrinsic responsiveness of microgels to external stimuli, making them highly relevant for the development of novel soft photonic materials. In this work, we used electrostatically assembled complexes to investigate their colloidal stability, an aspect that has received limited attention so far, despite its marked impact on the optical response of this hybrid system. We revealed a surprising re-entrant behavior of the plasmon coupling as a function of the NP-to-microgel number ratio $n$. By combining SAXS measurements of NP–NP spatial correlations with extinction spectroscopy, we rationalized the mechanisms underlying this phenomenology. To this aim, we established a quantitative relationship between plasmon coupling and the surface-to-surface distance $d$ between NPs adsorbed to microgels. Intriguingly, we found that this relationship is not universal, instead it branches in two distinct behaviors depending on the NP loading. At comparable interparticle distances, the plasmon coupling is significantly enhanced for intermediate surface coverages ($20\leq n\leq 100$), while for both low ($n=15$) and high ($n>100$) loadings it follows the same trend as a function of $d$ and remains systematically weaker with respect to intermediate $n$ values. 

We then demonstrated that this dual behavior originates from the aggregation of microgel–NPs complexes occurring only for intermediate $n$, as revealed by DLS measurements. In this regime, partial NP coverage results in oppositely charged patches on the surface of the complexes, promoting the formation of stable, finite-size clusters. Upon further increasing NP coverage, charge uniformity is progressively restored (with overcharging), electrostatic repulsion dominates, and the complexes regain colloidal stability. A clear correspondence is found between the aggregation phase-diagram and the deviation of the plasmon coupling from the trend of isolated microgel-NPs complexes.

On this basis, we propose that the splitting of the coupling-distance trends arises from the existence of two distinct plasmon coupling regimes. In colloidally stable microgel-NPs complexes, coupling is governed primarily by NP organization within the corona of individual microgels. In contrast, when the complexes aggregate, additional coupling emerges between NPs belonging to neighboring microgels, giving rise to a marked overall enhancement. This picture naturally explains the observed re-entrant behavior of the optical properties: when aggregation is suppressed, the coupling intensity decreases again.

Looking forward, we believe that the thermal reversibility of the plasmonic response, which we have shown to be largely recovered in the region of colloidal stability represents an important benchmark for designing fully reversible optical materials. Moreover, extending this study beyond the present case, it will be very interesting in the future to see whether these two coupling regimes also hold in different experimental conditions, varying for example the architecture of the polymer network through the fraction of crosslinker monomers, or NP size. In addition, the present case deals with microgel charges originating solely from the initiator, that are uniformly positive and localized in the outer corona, but there is the need to also explore how different spatial distributions of charges and the presence of oppositely charged groups within the polymer network affect NP adsorption and plasmon coupling.

\section{Experimental}
\subsection{Microgel synthesis}
Cationic microgels are synthesized by surfactant-free radical polymerization as detailed previously \cite{truzzolillo2018}, using NIPAM monomers (Sigma-Aldrich, MW = 113.16 Da), 5\% molar fraction of N,N'-Methylenebis(acrylamide) (BIS, Sigma-Aldrich, MW = 154.17 Da) as crosslinker and 1\% molar fraction of 2,2'-Azobis(2-methylpropionamidine) dihydrochloride (AIBA, Sigma-Aldrich, MW = 271.19 Da) as ionic initiator. Briefly, we dissolve 1.25 g NIPAM and 92 mg BIS in 148 mL of deionized water. Separately, 30 mg AIBA are dissolved in 2 mL of deionized water. The solution with the reagents is bubbled with argon for 30 minutes, we then heat it up to 70°C and add the initiator to start the reaction. After 6 h, the microgel dispersion obtained is cooled to room temperature and filtered through glass wool. To prevent bacterial growth, we add 2 mM \ce{NaN3}. The obtained microgels have hydrodynamic radius $R_H = $ 286 nm and number density $n_{mg}^{(0)} = 1.63 \times 10^{12}$ mL\textsuperscript{-1} (volume fraction $\varphi = 0.16$) at 20°C, as determined by dynamic light scattering (DLS) and viscosimetry measurements.

\subsection{Preparation of microgel--nanoparticles samples}
To prepare microgel--NPs samples, we use spherical gold NPs (Ted Pella) with nominal number density $n_\text{NP}^{(0)}=7.0\times10^{11}$ mL$^{-1}$. NPs are stabilized by a citrate capping that provides them with a negative charge, resulting in the electrophoretic mobility $\mu_e = -3.42 \times 10^{-8}$ m$^2$/Vs, and have a radius $a = 9.3$ nm, as previously evaluated \cite{brasili2025,brasili2023}. 
The microgel dispersion is diluted 250 times in 0.2 mM \ce{NaN3} to prevent bacterial growth. We verified by dynamic light scattering (DLS) measurements that this salt concentration is low enough to exclude any effect of ionic strength on the colloidal stability of the microgels or NPs. 
The NP dispersion is then either diluted with MilliQ water or concentrated by centrifugation to obtain the desired NP number densities for each sample. Finally, equal volumes of the microgel and NP dispersions are mixed to obtain samples with fixed $n_{mg} = 3.26 \times 10^9$ mL$^{-1}$ and varying NP-to-microgel number ratio $n = n_\text{NP}/n_{mg}$.

\subsection{Extinction spectroscopy}
Extinction spectra in the UV–Visible-NIR spectral range are acquired with a resolution of 0.1 nm, using a v-570 double ray spectrophotometer (Jasco, IT) equipped with a Peltier thermostated holder EHC-505. We measure spectra of the samples at varying temperature between 20 and 40°C; after every temperature change samples are kept thermalizing for at least 5 minutes and then spectra are acquired. All the spectra reported in the figures are normalized to the extinction at 400 nm, that corresponds to the gold interband transitions and is unaffected by particle size, shape, and environment~\cite{palmieri2022}. To quantify the modifications of the LSPR at varying temperature across microgels VPT, we use the wavelength $\lambda_\text{LSPR}$ of its maximum and the coupling degree, defined as the spectral weight of the region between 570 and 800 nm~\cite{brasili2025}, where coupled plasmon modes prevalently contribute to the extinction:
\begin{equation}
\label{eq:coupling}
\Delta\frac{A_C}{A_{tot}}=\frac{A_C(T)}{A_{tot}(T)}-\frac{A_C^{(NP)}}{A_{tot}^{(NP)}}\quad ,
\end{equation}
where $A(T)$ indicates the area underlying the spectrum acquired at temperature $T$; the pedices $tot$ and $C$ refer to the  overall spectral range and the region of coupled modes, respectively; $A_{tot}^{(NP)}$ and $A_C^{(NP)}$ are the areas computed in the same regions on the reference spectrum of non-interacting NPs. 

\subsection{Small angle X-ray scattering}
SAXS experiments are performed on samples filled in capillaries of 1.5 mm diameter and placed at 3 m sample-to-detector distance. 
The exposure time for acquisitions is set to 1 s and 14 scattering patterns are acquired for each sample. Scattering patterns are recorded at 12 keV using a two-dimensional EigerX 4-M detector (Dectris, Baden, Switzerland). This allows measurements in the range of $q$-vector between 0.002 and 0.38 \AA$^{-1}$, where $q$ is defined as $q=(4\pi/\lambda)\sin\theta$, $2\theta$ is the scattering angle, and $\lambda$ is the wavelength of the radiation. Scattering patterns of an empty capillary and of a capillary filled with water are recorded for normalization of the intensity to absolute units and background subtraction, respectively. Experiments are conducted at selected temperatures between 25°C and 40°C employing a Huber Ministat 125 thermostat. After each temperature change, samples are left to equilibrate for 5 minutes before measurements. The processing and averaging of the scattering patterns are performed by the software Foxtrot (SOLEIL software group and SWING beamline).

For a collection of particles, the scattered intensity $I(q)$ can be expressed in terms of the form factor $P(q)$ of single particles and of the structure factor $S(q)$ of the system as $I(q)=Nv^2\Delta\rho^2P(q)S(q)$, where $N$ and $v$ are the number density and the volume of the scattering particles, and $\Delta\rho$ is the contrast in electron density $\rho$ between particles and solvent. $S(q)$ is the interference introduced by interparticle correlations and can be expressed in terms of the Fourier transform of the pair correlation function $g(r)$ as:
\begin{equation}\label{eq:structure_factor}
S(q) = 1 + \rho \int_V g(r) e^{-i\vec{q}\cdot\vec{r}} d \vec{r} \quad .
\end{equation}
Since for a dilute system of non-interacting particles $S(q)=1$, we measure the form factor $P(q)$ of gold NPs on the stock solution \cite{capocefalo2022}. In microgel--NPs samples, the much higher electron density of gold relative to the polymer matrix ensures that, for the chosen acquisition times, the measured scattered intensity $I(q)$ originates exclusively from the NPs~\cite{suzuki2014}. The NP structure factor $S(q)$ is therefore obtained, for each sample, as $S(q)=I(q)/P(q)$. We analyze the $S(q)$ to extract the average surface-to-surface distance $d$ between nearest neighbor NPs. For microgel–NP complexes, two distinct $q$-regions can be identified, each one associated with structural correlations at different length scales~\cite{brasili2025}. At low $q$ (approximately below 0.1 nm$^{-1}$), the scattering captures microscopic features of the NP arrangement, e.g. reflecting the confinement of NPs within a spherical shell region in the microgel corona. At high $q$ (above 0.1 nm$^{-1}$), the range of interest here, the scattering probes nanoscale features of the local NP packing. We focus on the first peak in this high-$q$ range, from which we calculate the interparticle distance as:
\begin{equation}\label{eq:interparticle_distance}
d(n,T)=2\pi/q_p(n,T) - 2a \quad ,
\end{equation}
where $q_p(n,T)$ is the position of the peak for the sample $(n,T)$. We note that this peak identifies the dominant nearest-neighbor NP distance, irrespective of whether they are adsorbed to the same microgel or located on the surfaces of two approaching microgels.

\subsection{Hydrodynamic radius and electrophoretic mobility measurements}
Hydrodynamic radius $R_H$ and electrophoretic mobility $\mu_e$ of the suspended microgels are measured almost simultaneously by DLS and electrophoretic light scattering, respectively. We employed a NanoZetaSizer apparatus (Malvern Instruments LTD) equipped with a He-Ne laser (5 mW power, 633 nm wavelength). DLS intensity autocorrelation functions are measured at an angle of 173°, in quasi-backscattering geometry.  The correlograms are analyzed by means of the CONTIN algorithm \cite{provencher1982} to obtain the intensity weighted distributions of $R_H$. The electrophoretic mobility is obtained by phase analysis light scattering \cite{tscharnuter2001}, which allows accurate measurements of low-mobility samples. For soft, solvent-permeable colloids such as pNIPAM microgels in both swollen and collapsed states \cite{sbeih2019}, $\mu_e$ cannot be rigorously converted into a $\zeta$-potential or a surface charge density, since no sharp slipping plane exists when the flow penetrates the charged polymer network \cite{ohshima2009}. We therefore report and discuss the $\mu_e$ itself, which provides a robust, model-independent indicator of the net charge sign and its evolution with $n$.

Temperature trends are measured using ascending ramps between 25°C and 40°C with steps of 1°C. At each step, the sample is allowed to equilibrate for 5 minutes at the target temperature before performing measurements. The reported values of $R_H$ or $\mu_e$ and the associated errors are the mean and standard deviation of at least three independent measurements.

\subsection{Electron microscopy}
Transmission electron microscopy measurements were performed using a Tecnai G2 12 TWIN (Thermo Fisher Scientific) setup, operating at 120 kV and equipped with an electron energy loss filter (Biofilter, Gatan Inc.) and a slow-scan charge-coupled device camera (794 IF, Gatan Inc.). All the samples have been prepared by depositing 20 \textmu l of microgel-NPs dispersion on a 300 mesh copper grid covered by a thin amorphous carbon film. Following the protocol already used in a previous investigation \cite{truzzolillo2018}, samples have been deposited  both at room temperature and at 40 °C. The samples deposited at 25°C were stained by phosphotungstic acid, by adding 10 \textmu l of 2\% aqueous solution (with pH adjusted to 7.3 using 1 N \ce{NaOH}) to enhance image contrast and evidence the contour of microgels.

\section*{Acknowledgments}
The authors acknowledge SOLEIL for providing synchrotron radiation facilities under proposals n. 20191601 and n. 20240491 at the SWING beamline. F.Br., E.Z., F.Bo. and S.S. acknowledge financial support from INAIL, projects MicroMet (BRiC 2022, ID 16) and MicroPad (BRiC 2025, ID 66). F.Br., E.Z. and S.S. also acknowledge support from ERC POC project MICROSENS (grant agreement n. 101157420). D.T and E.C. acknowledge financial support from the Agence Nationale de la Recherche (Grant ANR-20-CE06-0030-01; THELECTRA).

\section*{Data Availability Statement}
The experimental data supporting the findings of this work are openly available on Zenodo at doi:10.5281/zenodo.20119924.

\bibliographystyle{unsrt}
\bibliography{references}

\clearpage
\setcounter{section}{0}
\setcounter{figure}{0}
\setcounter{table}{0}
\setcounter{equation}{0}
\renewcommand{\theequation}{S\arabic{equation}}
\renewcommand{\thefigure}{S\arabic{figure}}
\renewcommand{\thetable}{S\arabic{table}}
\renewcommand{\thesection}{S\arabic{section}}

\makeatletter
\def\@seccntformat#1{\csname the#1\endcsname.\quad}
\makeatother

\begin{center} {\huge Supporting Information} \end{center}
\vspace{1em}

\section{Additional extinction spectroscopy and SAXS data}
\begin{figure}[b!]
\centering
\includegraphics[width=0.9\linewidth]{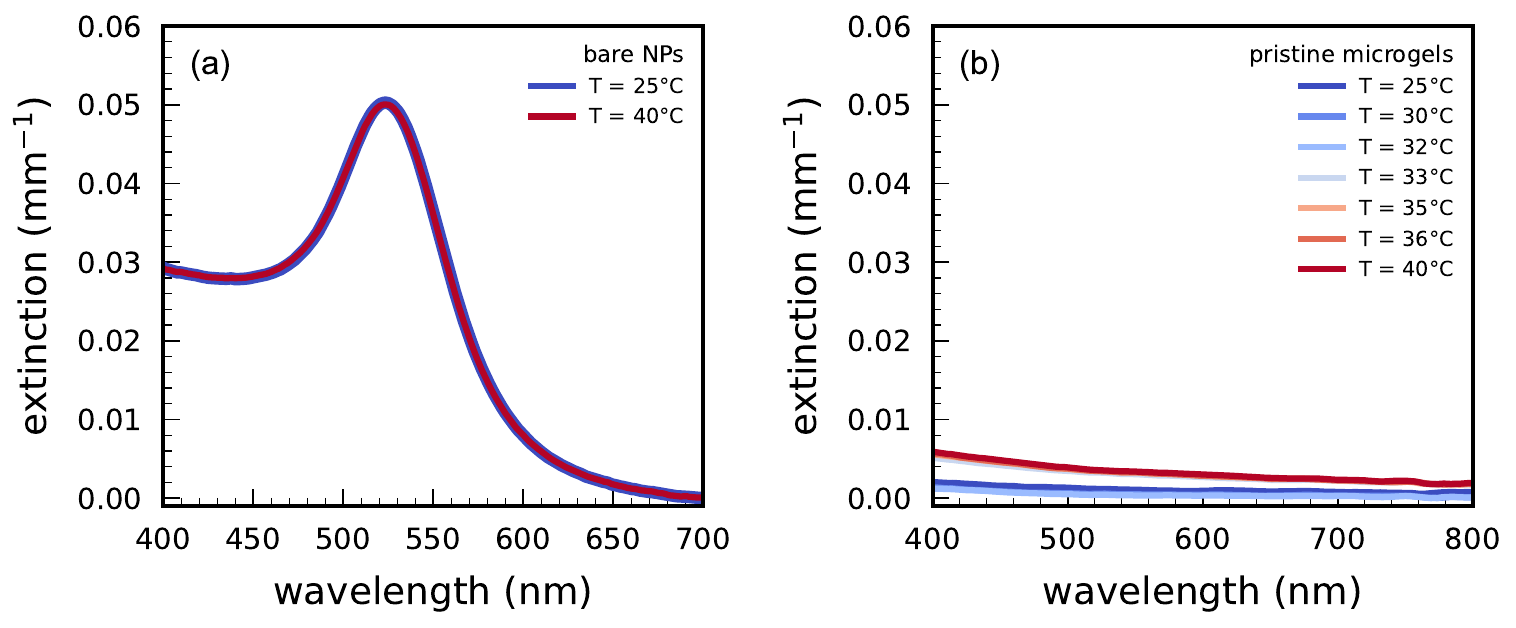}
\caption{(a) Extinction spectra of stock NPs, at the same NP concentration of the microgel-NPs sample with $n=150$, at 25°C and 40°C. (b) Extinction spectra of pristine microgels at varying temperature across the VPT.}\label{fig:figS1_refs}
\end{figure}
Reference extinction spectra of the two components of the hybrid microgel-NPs complexes, namely gold nanoparticles (NPs, 20 nm diameter) and pristine microgel dispersions, are shown in Figs.~\ref{fig:figS1_refs}a and \ref{fig:figS1_refs}b, respectively, at different temperatures across the microgel volume phase transition (VPT). The sample numbers densities are $n_\text{NP} = 4.89 \times 10^{11}$ mL$^{-1}$ and $n_{mg} = 3.26 \times 10^9$ mL$^{-1}$ for NPs and microgels, respectively, the same of the microgel--NPs sample with the NP-to-microgel number ratio $n=150$.  To directly compare the two datasets, the spectra are reported in absolute extinction units rather than being normalized to the extinction at 400 nm, using the same scale. 
The NP spectrum exhibits the characteristic localized surface plasmon resonance (LSPR) peak at $\lambda_\text{LSPR}=525$ nm and remains unchanged between 25°C and 40°C. In contrast, the microgel spectra evolve with temperature as the particles collapse, becoming smaller and denser across the VPT, leading to an increasing turbidity.\cite{otten2024} 
Comparison of the two spectra shows that, even above the VPT where the microgel contribution is largest, the maximum extinction of the pristine microgel dispersion remains at least fivefold smaller than the LSPR peak of the NP dispersion. The contribution of the microgels to the extinction spectra of the hybrid complexes can therefore be considered negligible. Moreover, even in the lowest-$n$ samples, where the microgel contribution may become more appreciable, its monotonically decreasing spectral profile would only lead to a slight underestimation of the LSPR shift $\Delta\lambda_\text{LSPR}$ and of the coupling degree $\Delta (A_C/A_{tot})$ at high temperatures. A thorough analysis of these aspects is reported in Section \ref{sec:S3_corrections}.

\begin{figure}[h]
\centering
\includegraphics[width=0.6\linewidth]{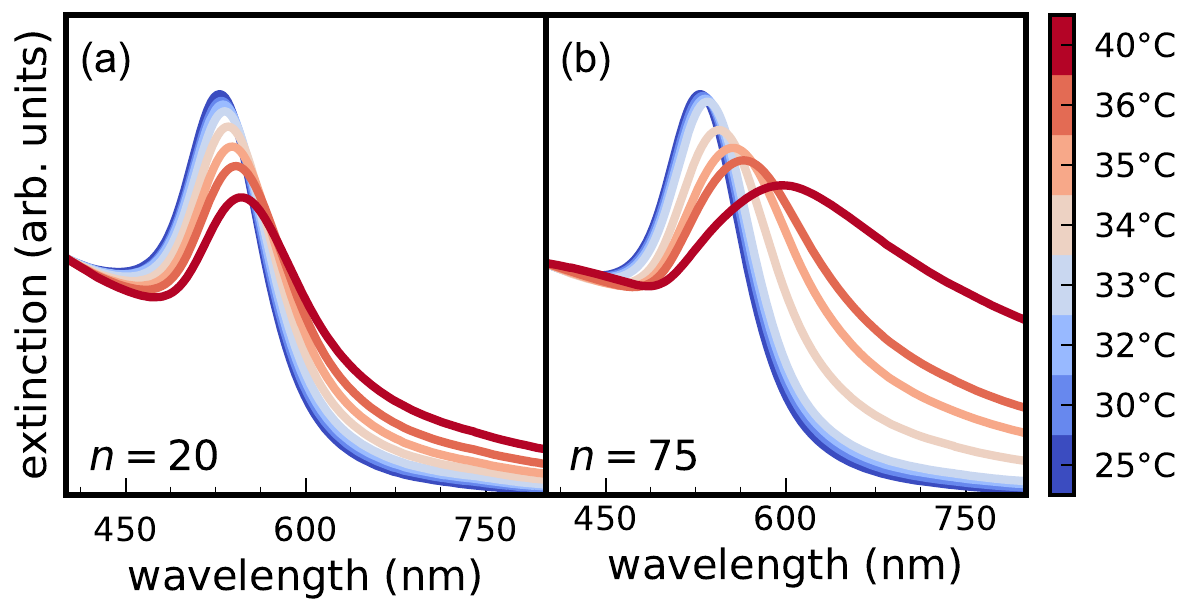}
\caption{Extinction spectra of microgel-NPs at different temperatures for $n=20$ (a) and $n=75$ (b).}\label{fig:figS2_spectra}
\end{figure}

\begin{figure}[h]
\centering
\includegraphics[width=0.55\linewidth]{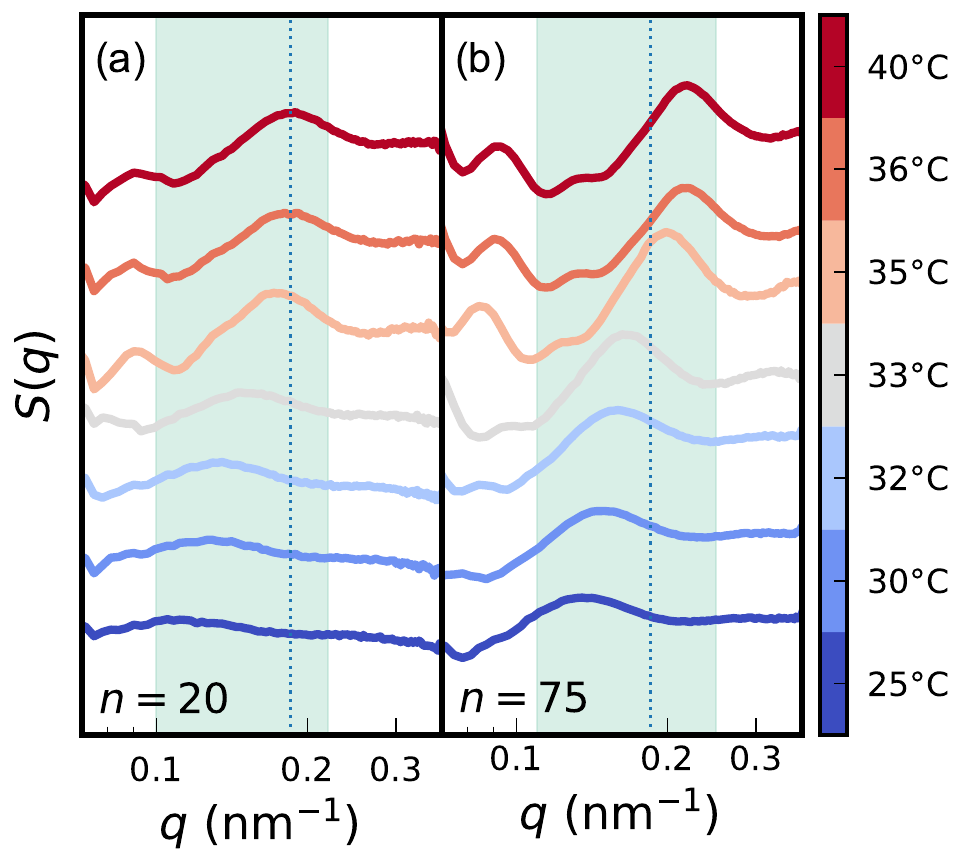}
\caption{Structure factors of the microgel-NPs samples at different temperatures for $n=20$ (a) and $n=75$ (b). The light green color highlights the peaks corresponding to the distance between NPs.}\label{fig:figS3_saxs}
\end{figure}

Additional extinction spectra and the structure factors $S(q)$ of the microgel--NPs samples are reported in Fig.~\ref{fig:figS2_spectra} and Fig.~\ref{fig:figS3_saxs}, respectively, Specifically, we show in the figures the datasets at varying temperature for $n=20$ and $n=75$ samples. The corresponding datasets for the $n=150$ sample were previously published in ref.~\citenum{brasili2025} and are also available through the Zenodo repository (doi:10.5281/zenodo.20042415).

\section{Charge-patch asymmetry accounts for the \texorpdfstring{$n$}{n}-offset between charge inversion and aggregation maximum} \label{sec:S2_overcharging}
The electrophoretic mobility $\mu_e$ of the microgel-NPs complexes (Fig.~\ref{fig:fig5_dls}b) provides a robust, model-in\-de\-pend\-ent measurement of the sign of the net charge and of the loading $n$ at which it reverses. We note, however, that electro-osmotic contributions to the electrophoretic force can shift the mobility reversal with respect to the point of zero net charge, particularly when ion-ion correlations in the Debye layer are strong~\cite{stout2014,sennato2021}. Since ion-ion correlations are enhanced by high adsorption energies -- which are expected to be larger for the collapsed microgels due to their higher surface charge density -- the condition $\mu_e = 0$ is a more reliable marker for the isoelectric point at low temperature, where the microgels are swollen and correlations are weaker. We therefore use the charge inversion observed at 25°C, $n \simeq 30$, as our best estimate of the charge inversion. Based on this, we derive an order-of-magnitude estimate of the charge inhomogeneity between NP-covered and bare regions of the particles' surface.

We model the surface of a complex as composed of two types of patches: a fraction $\theta$ covered by adsorbed NPs, carrying an effective negative charge density $-\sigma_\text{NP}$, and the complementary bare-microgel fraction $1-\theta$, carrying a positive charge density $+\sigma_{\mu\text{gel}}$. Within this two-patch description, $\theta$ can be evaluated as the projected-area coverage of the $n_{ads}$ NPs adsorbed to the microgel as: 
$$\theta = n_{ads}\frac{\pi a^2}{4\pi R_{\mu\text{gel}}^2}  \quad ,$$ 
where $R_{\mu\text{gel}}$ is the radius of the adsorbing microgel and $a = 9.3$ nm is the NP radius, and the mean net charge density is given by:
$$
<\sigma>\;=\,(1-\theta)\,\sigma_{\mu\text{gel}} - \theta\,\sigma_\text{NP} \quad ,
$$
which vanishes at a coverage $\theta_0 = \frac{\sigma_{\mu\text{gel}}}{\sigma_\text{NP}+\sigma_{\mu\text{gel}}}$, so that:

$$\frac{\sigma_\text{NP}}{\sigma_{\mu\text{gel}}} = \frac{1-\theta_0}{\theta_0} \quad .$$

We now focus on the microgel at $T = 40$°C ($R_{\mu\text{gel}}\approx 100\div105$ nm), since aggregation occurs only in the collapsed state. As established in ref.~\citenum{brasili2023} for less charged microgels and thus still valid for the slightly more charged microgels used here, for $n=30$ essentially all NPs are adsorbed to the microgels ($n_{ads}\simeq n$) and, pushed outwards during microgel collapse without significant desorption, at 40°C they are all located on the external surface of the hybrid complex \cite{brasili2025}. We can therefore estimate $\theta_0\approx 6\%$ and, consequently, $\sigma_\text{NP}/\sigma_{\mu\text{gel}}\approx15$.

Although this estimate gives only an order of magnitude and rests on simplifying assumptions (two discrete patch populations, uniform charge densities within each, and a projected-area coverage), it clearly highlights that charge reversal is achieved at a very low surface coverage (see also electron microscopy in Fig.~\ref{fig:fig5_dls}d), implying that the NP-covered patches are far more densely charged than the bare microgel surface. Consistently, the coverage remains low over the entire range of $n$ in which aggregation occurs ($\theta<25\%$ for $n\leq100$), so that the surface never becomes uniformly charged and the system always resides in the charge-patch regime.

The strong asymmetry $\sigma_\text{NP}>>\sigma_{\mu\text{gel}}$ also clarifies why the maximum aggregate size is reached at a loading well beyond the charge-inversion point (compare Figs.~\ref{fig:fig5_dls}a and \ref{fig:fig5_dls}b). Because the NP-covered patches are much more densely charged than the bare microgel surface, net charge neutrality is reached at a very low coverage. In contrast, the attractive contribution to interactions arising from charge inhomogeneity scales geometrically as $\theta(1-\theta)$\cite{brasili2020}, becoming stronger with $n$ well past $\theta_0$. 
The aggregation maximum is therefore shifted to higher loadings respect to net neutrality, where the inhomogeneity is substantial while the (now overcharged) average surface charge density is still moderate; at higher $n$, the growing surface charge restabilizes the complexes through electrostatic repulsion between facing portions of the surfaces of approaching particles. Had the two patch densities been comparable ($\sigma_\text{NP}\approx\sigma_{\mu\text{gel}}$), net neutrality and maximum inhomogeneity would both fall near $\theta=0.5$, and the two features would nearly coincide.

\section{Effect of microgel collapse and aggregation on the extinction spectra} \label{sec:S3_corrections}
\begin{figure}[t!]
\centering
\includegraphics[width=\linewidth]{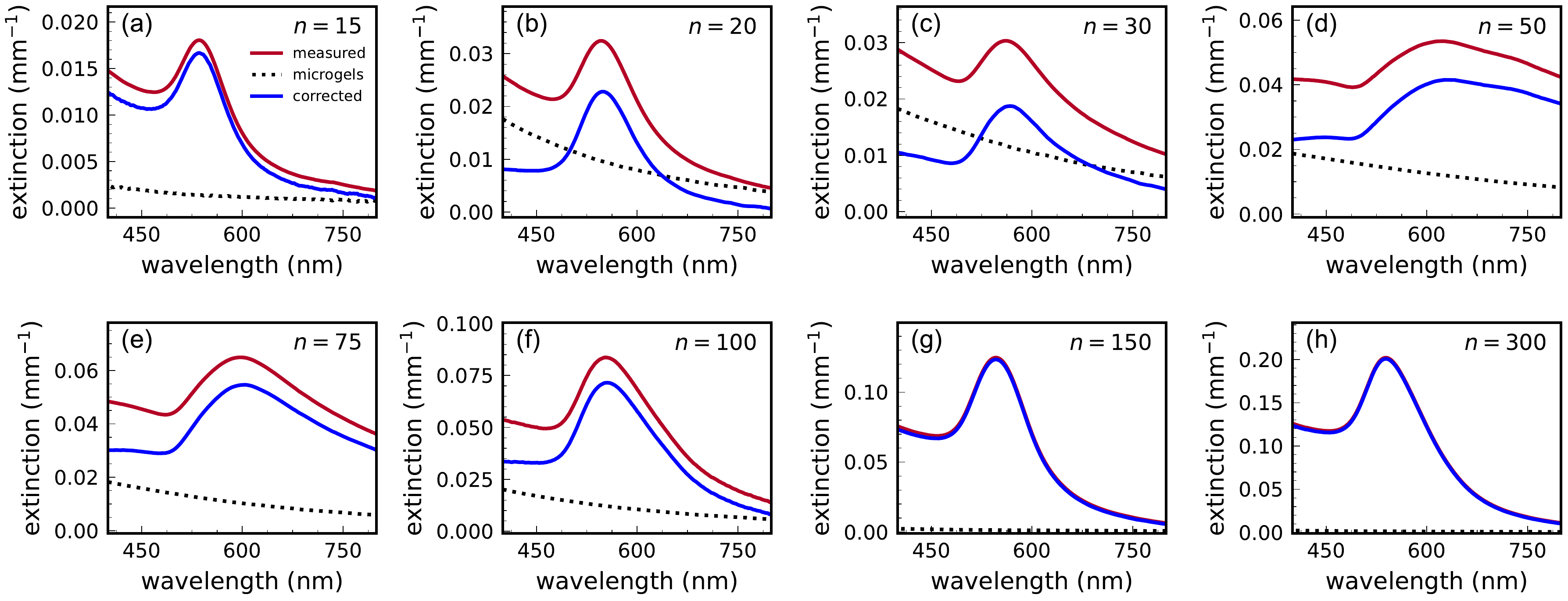}
\caption{Correction of the extinction spectra for the scattering contribution of NP-free microgel aggregates, for each of the NP-to-microgel number ratios $n$ investigated in this work. In each panel, the red solid line is the measured extinction spectrum of the microgel–NP suspension at 40°C; the black dotted line is the extinction spectrum of a NP-free microgel aggregate sample at 40°C, obtained by salt-induced aggregation and size-matched to the corresponding microgel–NP suspension (same microgel number concentration); the blue solid line is the corrected spectrum, obtained by subtracting the microgel aggregate spectrum from the measured one.}\label{fig:figS4_spectra_corr}
\end{figure}

As discussed in the main text, the microgel–NP suspensions may contain mesoscale aggregates at temperatures above the microgel VPT, depending on the NP-to-microgel number ratio $n$ (Fig.~\ref{fig:fig5_dls}). Because large colloidal aggregates scatter light with higher intensity relative to a non-aggregated dispersion, we assessed whether this contribution could partly account for the observed modifications of the extinction spectra and the apparent increase in the coupling degree $\Delta (A_C/A_{tot})$.
\begin{figure}[b!]
\centering
\includegraphics[width=0.75\linewidth]{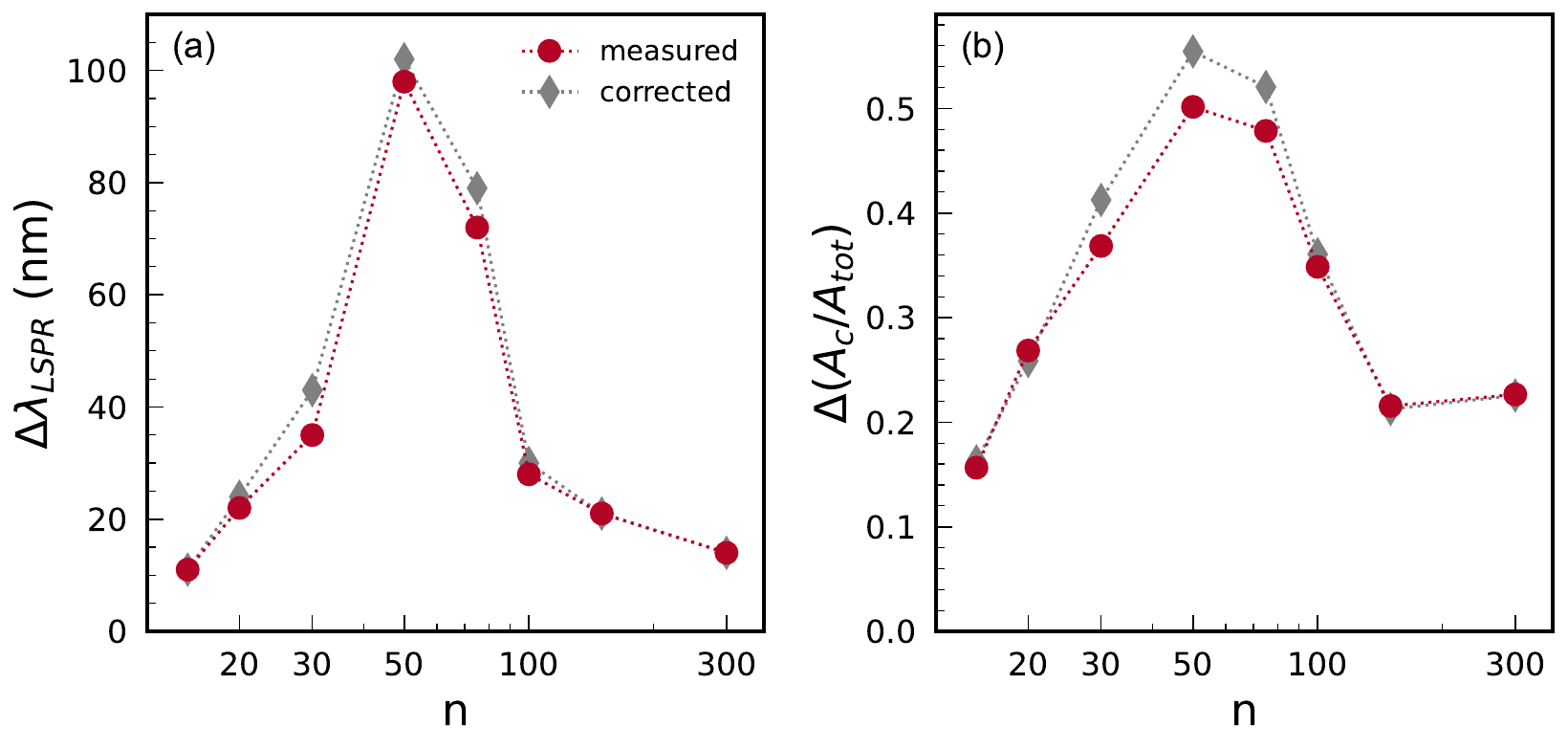}
\caption{Effect of the aggregate scattering correction on the plasmonic observables at 40°C. (a) LSPR peak redshift $\Delta\lambda_\text{LSPR}$ and (b) degree of plasmon coupling $\Delta (A_C/A_{tot})$ as a function of the NP-to-microgel number ratio $n$, before (red circles) and after (grey diamonds) subtraction of the size-matched NP-free microgel aggregate spectra (see Fig.~\ref{fig:figS4_spectra_corr}). In both panels, the corrected values are systematically larger than the uncorrected ones. The re-entrant dependence on $n$, with a maximum at $n=50$, is fully preserved after the correction.}\label{fig:figS5_coupling_corr}
\end{figure}

To obtain a reference for the scattering background, we induced aggregation in pristine (NP-free) microgel dispersions by adding \ce{NaCl} at concentrations between 5 and 50 mM and equilibrating at 40°C, where aggregation, and hence the potential scattering artefact, is most pronounced. In all reference samples, the microgel number concentration was kept identical to that of the corresponding microgel–NP suspensions, so that differences in the extinction spectra reflect only the presence of the NPs and their plasmonic response. DLS measurements confirmed that this procedure yields aggregates with hydrodynamic radii spanning the range observed in the microgel–NP samples. For each value of $n$, we identified the NP-free aggregate sample with the closest matching $R_H$ and used its extinction spectrum as a proxy for the non-plasmonic scattering background. For the three values of $n$ at which the microgel–NP suspensions do not exhibit aggregation ($n=15$, $n=150$ and $n=300$), the reference spectrum was taken as that of the pristine microgel dispersion at 40°C in the absence of added salt (Fig.~\ref{fig:figS1_refs}). Fig.~\ref{fig:figS4_spectra_corr} shows, for each value of $n$ investigated, three spectra: (i) the extinction spectrum of the microgel–NP suspension at 40°C, (ii) the extinction spectrum of the size-matched NP-free microgel aggregate, and (iii) the corrected spectrum obtained by subtracting (ii) from (i). Unlike the spectra shown in the main text, which are normalized to the extinction at 400 nm, here all spectra are reported in absolute extinction units, since the background subtraction requires that the two spectra being compared are on the same scale. In all cases, the reference aggregate spectra exhibit the expected featureless, monotonically decreasing profile characteristic of dielectric particles. After subtraction, the LSPR peak and the long-wavelength coupled-plasmon shoulder remain clearly resolved, confirming that the spectral features attributed to plasmon coupling in the main text are not an artefact of aggregate scattering. We note that this subtraction procedure is inherently approximate. NP-free microgel aggregates and microgel–NP aggregates with the same $R_H$ may differ in internal structure, compactness, and fractal dimension, and therefore need not have identical scattering cross-sections per unit volume. The corrected spectra should accordingly be regarded as estimates of the plasmonic contribution, rather than exact decompositions.

Fig.~\ref{fig:figS5_coupling_corr} compares the $n$-dependence of $\Delta\lambda_\text{LSPR}$ and of $\Delta (A_C/A_{tot})$ before and after the background correction, evaluated at 40°C. In both cases, the corrected values are systematically larger than the uncorrected ones. This outcome is a direct consequence of the monotonically decreasing shape of the aggregate scattering spectrum: removing a background that is relatively larger at shorter wavelengths increases the relative weight of the long-wavelength shoulder and shifts the apparent peak position to the red. The uncorrected values reported in the main text are therefore conservative underestimates of the true degree of plasmon coupling. Most importantly, the re-entrant dependence of both $\Delta\lambda_\text{LSPR}$ and $\Delta (A_C/A_{tot})$ on $n$ -- the central result of this work -- is fully preserved after the correction, and indeed appears more pronounced. 

\section{Assessment of thermal reversibility} \label{sec:S4_reversibility}
\begin{figure}[t]
\centering
\includegraphics[width=\linewidth]{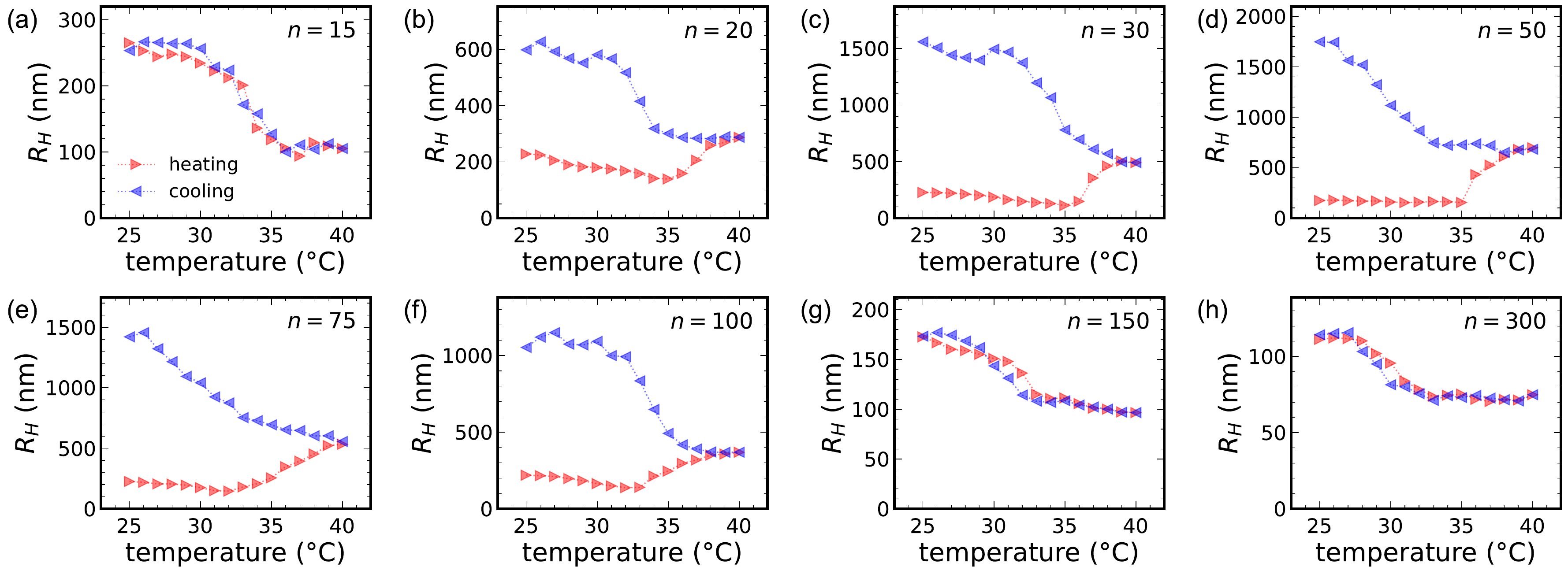}
\caption{Thermal reversibility of the hydrodynamic radius. $R_H$ is shown as a function of temperature during heating (25°C $\rightarrow$ 40°C, light red triangles) and cooling (40°C $\rightarrow$ 25°C, light blue triangles) for each of the NP-to-microgel number ratios $n$ investigated in this work. For $n = 15$, $150$, and $300$, the two branches overlap throughout, indicating reversible swelling and deswelling. For intermediate $n$, the cooling branch reaches values of $R_H$ at 25°C far exceeding the initial ones, demonstrating that the formed aggregates do not redisperse.}\label{fig:figS6_rev_aggregation}
\end{figure}
\begin{figure}[t!]
\centering
\includegraphics[width=\linewidth]{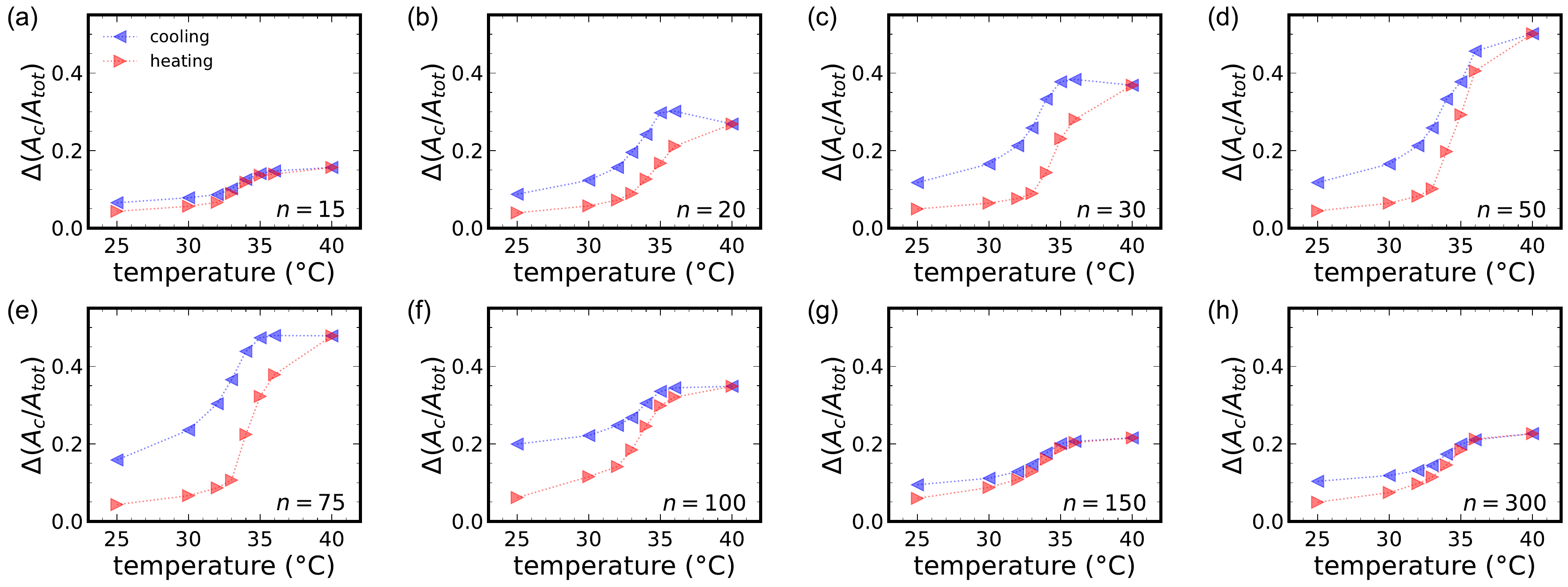}
\caption{Thermal reversibility of the degree of plasmon coupling $\Delta(A_C/A_\mathrm{tot})$. Data are shown during heating (light red triangles) and cooling (light blue triangles) for each of the NP-to-microgel number ratios $n$ investigated in this work. For non-aggregating samples ($n = 15$, $150$, and $300$), the two branches closely overlap over most of the temperature range, with a small residual offset at 25°C after cooling. For aggregating samples ($20\leq n\leq100$), the coupling degree remains elevated upon cooling, reflecting the persistence of strongly coupled NP configurations within the aggregates.}\label{fig:figS7_rev_redshift}
\vspace{1cm}
\includegraphics[width=\linewidth]{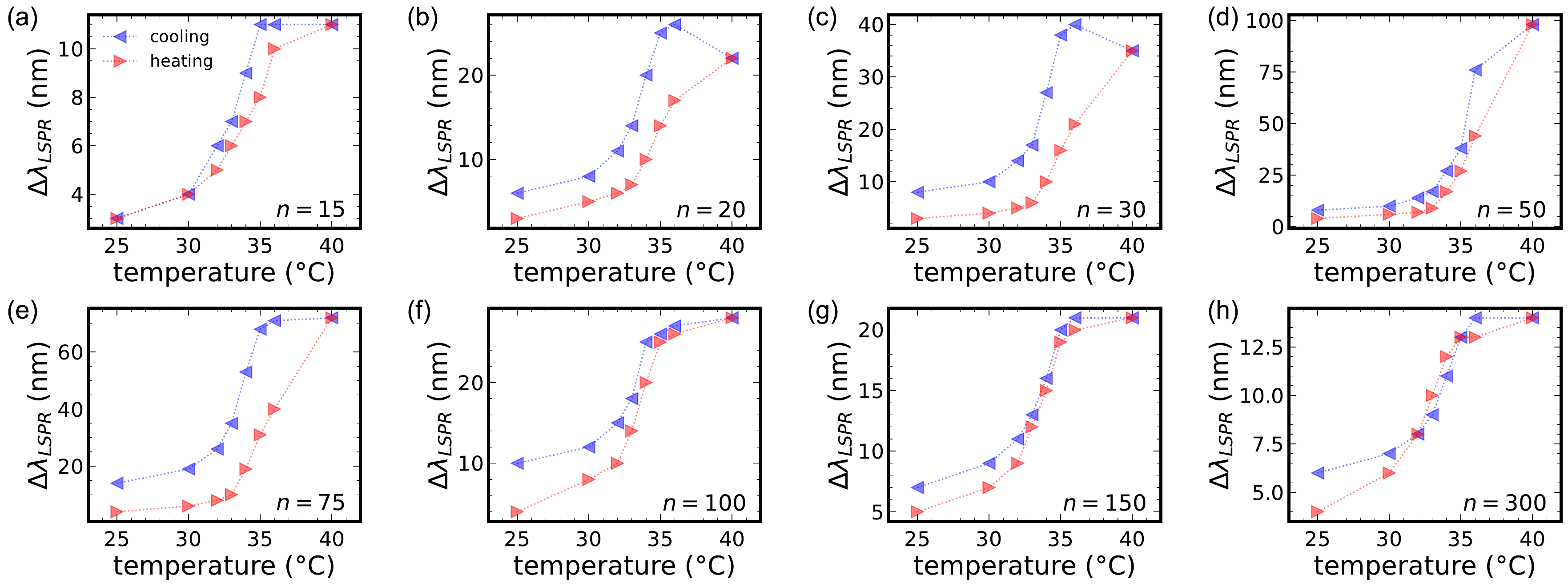}
\caption{Thermal reversibility of the LSPR peak shift $\Delta\lambda_\mathrm{LSPR}$. Data are shown during heating (light red triangles) and cooling (light blue triangles) for each of the NP-to-microgel number ratios $n$ investigated in this work. The behaviour mirrors that of $\Delta(A_C/A_\mathrm{tot})$ (Fig.~\ref{fig:figS7_rev_redshift}): largely reversible for non-aggregating samples with a small residual offset at 25°C, and strongly irreversible for aggregating samples.}\label{fig:figS8_rev_coupling}
\vspace{1.1cm}
\end{figure}

\paragraph{Colloidal stability.} Fig.~\ref{fig:figS6_rev_aggregation} shows $R_H$ as a function of temperature during heating (light red triangles) and cooling (light blue triangles). For $n = 15$, $150$, and $300$, the two branches overlap closely across the entire temperature range, confirming that the VPT of the NP-decorated microgels is thermally reversible and that colloidal stability is maintained throughout the cycle. For intermediate $n$ ($20\leq n\leq100$), the cooling branch leads to values of $R_H$ at 25°C far exceeding the initial ones, demonstrating that the charge-patch-driven aggregation is irreversible. We note that this behaviour differs from that observed for microgels complexed with linear polyelectrolyte, where charge-patch-driven clusters can redisperse upon cooling~\cite{truzzolillo2018}. Whether the irreversibility observed here is due to the higher surface charge density of the NPs, to their larger size, to their different chemical nature, or to a combination of these factors, remains an open question that needs further investigation.

\paragraph{Plasmonic observables.} Figs.~\ref{fig:figS7_rev_redshift} and \ref{fig:figS8_rev_coupling} show $\Delta(A_C/A_\mathrm{tot})$ and $\Delta\lambda_\mathrm{LSPR}$, respectively, as a function of temperature during heating and cooling. For the aggregating samples ($20\leq n\leq100$), both observables display pronounced hysteresis consistent with the irreversible aggregation observed by DLS: the strongly coupled configurations induced by aggregate formation persist upon cooling. For the non-aggregating samples ($n = 15$, $150$, and $300$), the cooling branches closely follow the heating ones over most of the temperature range, but a small residual offset at 25°C indicates that the optical response is not perfectly reversible.
\begin{figure}[t!]
\centering
\includegraphics[width=0.95\linewidth]{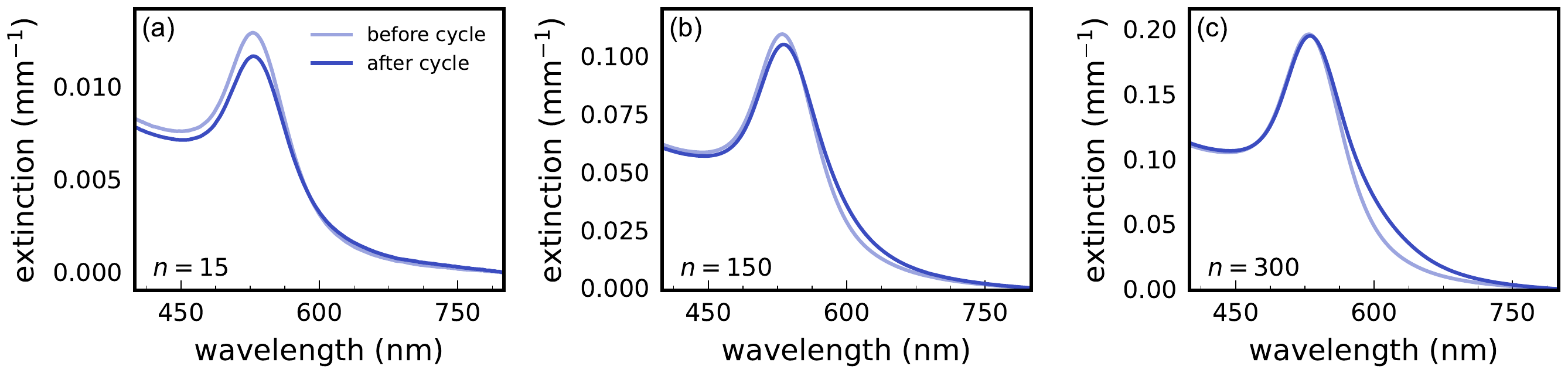}
\caption{Extinction spectra at 25°C before (light blue) and after (dark blue) a complete heating–cooling cycle for non-aggregating samples: (a) $n = 15$, (b) $n = 150$, and (c) $n = 300$.}\label{fig:figS9_spectra_comparison}
\end{figure}
To clarify the origin of the residual offset in the plasmonic observables at 25°C for the non-aggregating samples ($n = 15$, $150$, and $300$), we compare in Fig.~\ref{fig:figS9_spectra_comparison} the extinction spectra recorded at 25°C before and after the thermal cycle. We could identify two regimes with different behavior. At $n = 15$ (panel a), the post-cycle spectrum shows a slight reduction of the LSPR peak height with no appreciable increase of the long-wavelength shoulder ($\lambda > 570$ nm). This indicates that the increase in $\Delta(A_C/A_\mathrm{tot})$ after the cycle does not arise from near-field coupling between NPs, but rather from a modification of the local dielectric environment of the NPs. We tentatively attribute this to polymer chains from the microgel network that adsorb onto the NP surface during collapse and remain attached after re-swelling, increasing the effective refractive index sensed by the NPs and broadening their LSPR. At $n = 300$ (panel c), and to a lesser extent at $n = 150$ (panel b), the post-cycle spectrum shows instead a clear enhancement of extinction at $\lambda > 570$ nm, with the emergence of a coupled-plasmon shoulder absent before the cycle. This is a direct spectroscopic signature of residual near-field coupling between NPs that have not fully re-separated upon microgel re-swelling. In the high-$n$ regime, the large number of NPs per microgel causes close NP–NP contacts to form during collapse; polymer-mediated interparticle attractions can then stabilize these configurations, preventing complete re-separation upon re-swelling.

\end{document}